\documentclass[final,5p,times,twocolumn]{elsarticle}
\usepackage{amssymb}
\usepackage{algorithm}
\usepackage{algorithmic}
\usepackage{dsfont}
\usepackage{balance}
\usepackage{csquotes}
\usepackage{graphicx}
\usepackage{animate}
\usepackage{subfig}
\usepackage{tikz}
\usepackage{orcidlink}
\usepackage{amsmath}
\usepackage{hyperref}
\usepackage[normalem]{ulem}
\usepackage{todonotes}
\usepackage{setspace}
\usepackage{xcolor}
\usepackage{draftwatermark}
\SetWatermarkText{DRAFT}
\SetWatermarkScale{5.0}

\begin{document}
\begin{frontmatter}
%
% Title --------------------------------------------------------------------------
\title{Investigating the price determinants of the European Emission Trading System:\\ a non-parametric approach}
%
% Authors ------------------------------------------------------------------------
\author[inst1]{Cristiano Salvagnin\corref{cor1} \orcidlink{0000-0002-3348-0376}}
\ead{c.salvagnin@unibs.it}
\cortext[cor1]{Corresponding authors}
\author[inst2]{Aldo Glielmo\corref{cor1} \orcidlink{0000-0002-4737-2878}}
\ead{aldo.glielmo@bancaditalia.it}
\author[inst3]{Maria Elena De Giuli \orcidlink{0000-0002-5221-0005}}
\author[inst4,inst5]{Antonietta Mira \orcidlink{0000-0002-5609-7935}}
%
% Affiliations -------------------------------------------------------------------
\affiliation[inst1]{organization={Department of Economics and Management, University of Brescia},
            addressline={Via S. Faustino 74/b}, 
            city={Brescia},
            postcode={25122},
            country={Italy}}
\affiliation[inst2]{organization={Applied Research Team, Directorate General for IT, Banca d'Italia},
            country={Italy}}
\affiliation[inst3]{organization={Department of Economics and Management, University of Pavia},
            addressline={Via S. Felice Al Monastero, 5}, 
            city={Pavia},
            postcode={27100},
            country={Italy}}
\affiliation[inst4]{organization={Faculty of Economics, Euler Institute, Università della Svizzera italiana (USI)},
            addressline={
            Via Giuseppe Buffi, 13}, 
            city={Lugano},
            postcode={6900},
            country={Switzerland}}
\affiliation[inst5]{organization={Department of Science and High Technology, University of Insubria},
            addressline={Via Valleggio, 11}, 
            city={Como},
            postcode={22100},
            country={Italy}}
\singlespacing
%
% ABSTACT ------------------------------------------------------------------------
\begin{abstract}
    The European carbon market plays a pivotal role in the European Union's ambitious target of achieving carbon neutrality by 2050. Understanding the intricacies of factors influencing European Union Emission Trading System (EU ETS) market prices is paramount for effective policy making and strategy implementation. We propose the use of the Information Imbalance, a recently introduced non-parametric measure quantifying the degree to which a set of variables is informative with respect to another one, to study the relationships among macroeconomic, economic, uncertainty, and energy variables concerning EU ETS prices. Our analysis shows that in Phase 3 commodity related variables such as the ERIX index are the most informative to explain the behaviour of the EU ETS market price. Transitioning to Phase 4, financial fluctuations take centre stage, with the uncertainty in the EUR/CHF exchange rate emerging as a crucial determinant. These results reflect the disruptive impacts of the COVID-19 pandemic and the energy crisis in reshaping the importance of the different variables. Beyond variable analysis, we also propose to leverage the Information Imbalance to address the problem of mixed-frequency forecasting, and we identify the weekly time scale as the most informative for predicting the EU ETS price. Finally, we show how the Information Imbalance can be effectively combined with Gaussian Process regression for efficient nowcasting and forecasting using very small sets of highly informative predictors.
\end{abstract}
%
% Keywords -----------------------------------------------------------------------
\begin{keyword}
    EU ETS \sep Information Imbalance \sep Gaussian Processes \sep Feature selection \sep Mixed frequency data \sep Forecasting  % \sep Mixed-data sampling \sep Time scale selection
    %CRI Abbiamo solo 6 slot per le Keyword, adesso ne abbiamo 8. Proprongo di togliere Time scale selection, che forse dice poco e aggiungere una tra Mixed-data samplig e Mixed frequency data

% JEL codes ----------------------------------------------------------------------
    \JEL Q56 \sep F18 \sep Q58 \sep D40 \sep D80 \sep D47
\end{keyword}
\end{frontmatter}
%
% Section 1: Introduction --------------------------------------------------------
\onehalfspacing
\section{Introduction}
\label{sec:introduction}
    The European Union Emission Trading System (EU ETS) is a key component of the European Union's effort to combat climate change and reduce greenhouse gas (GHG) emissions. The EU ETS sets a gradual limit on GHG emissions in key sectors of the economy, mainly the energy sector and industry. The European Union issues emission permits, which authorise the discharges of one tonne of carbon dioxide or its equivalent. Participating companies can buy and sell these allowances on the ETS market. If a company emits less than its allowances, it can sell the surplus; conversely, if it exceeds its allowances, it must buy more or face financial penalties. This market-based approach offers flexibility in reducing emissions by providing a financial incentive for companies that reduce emissions below their allocated allowances, while punishing those that exceed these limits. Companies can decide whether to invest in emission reduction initiatives or buy additional allowances from the market, depending on which option is more advantageous from an economic and sustainability perspective. The system is periodically reviewed to reduce emission limits (also called \textit{cap}). \citep{bersani2022ets} note that the equilibrium price of certificates is mean-reverting, where the fundamental mechanism is the excess or shortfall of emissions over the cap set by the regulator, which can be inferred in practice through the number of certificates distributed.
    
    The EU ETS is considered one of the world's largest and most established emissions trading systems. In operation since 2005, it aims to promote innovation, stimulate investment in clean technologies and contribute to the EU's overall climate goals by reducing greenhouse gas emissions in an economically sustainable manner, in accordance with the Kyoto Protocol and the Paris Agreement.
%
%   Subsection 1.1: Phase 3 and Phase 4 ------------------------------------------
    \subsection{Phase 3 and Phase 4}
        \begin{figure}[htp]
            \centering
            \includegraphics[width=0.95\columnwidth]{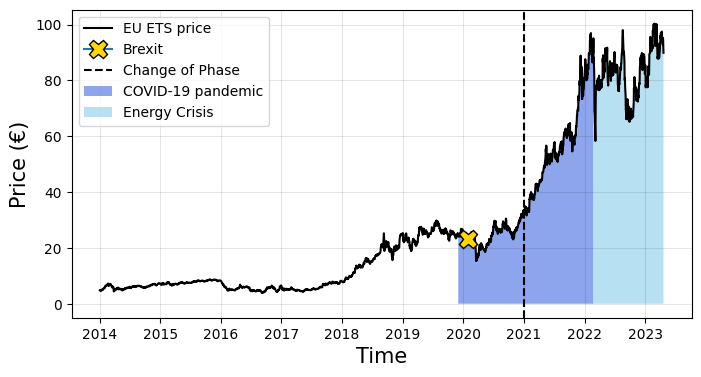}
            \caption{\textbf{EUA price.}
            Evolution of the EUA price from January 2014 to April 2023.
            The dashed line shows the change between Phases 3 and 4.
            Three major events are also displayed: Brexit, the COVID-19 pandemic and the energy crisis due to the Ukrainian conflict.}
            \label{fig:1}
        \end{figure}

        The EU ETS is organised into distinct phases, each defined by specific rules, objectives, and regulations. These phases serve as a road map for the progressive establishment and development of the system. Currently, the EU ETS has traversed three primary phases, while a fourth phase is presently underway; in this work we consider both Phase 3 and Phase 4. In Phase 3 of the EU Emissions Trading Scheme, a deliberate decrease in the overall cap on emissions was established to encourage incremental reductions in emissions. Introduced in Phase 3, the Market Stability Reserve (MSR) was intended to eliminate the effect of excess allowances on the market by reducing the number of allowances allocated to this reserve in order to ensure greater market price stability for permits. During Phase 3, the system included both free allocation and auctioning of allowances. Certain industries, particularly those prone to high carbon emissions due to the characteristics of the industry itself, received a free allocation of allowances in order to mitigate the risk of relocation to regions with less stringent emission regulations. Example sectors include power generation, heavy industry and aviation. In Phase 4, the EU has committed to more ambitious emission reduction targets by 2030. The main target is to reduce greenhouse gas emissions by at least 40\% compared to 1990 levels. As of 2022, the start of Phase 4, the Linear Reduction Factor (LRF) was increased to 2.2\% per year, leading to a faster decrease in the emissions cap than in Phase 3. In addition, there was an increase in the proportion of allowances provided through auctions, with the aim of phasing out free allocation in specific sectors. This change mainly affects sectors that benefited during Phase 3, while maintaining free allocation for strategic but carbon-dependent industries \citep{euets}.
%
%   Subsection 1.2: Motivation ---------------------------------------------------
    \subsection{Motivation}
        In recent times, energy markets have been closely monitored by governments, investors and society at large. In Europe in particular, a series of extraordinary events such as the COVID-19 pandemic and the development of the conflict between Ukraine and Russia caused an increase in the volatility of energy prices, resulting in an energy crisis with a consequent increase in commodity market prices. As shown in Figure \ref{fig:1}, the impact of these two crises, health and energy, manifested itself mainly in Phase 4, leading to a significant increase and variability of permit prices compared to the previous phase. High energy costs directly impact the operational expenses of EU ETS participants, potentially obstructing their compliance with emission reduction goals and impeding progress towards mandated targets. Furthermore, the energy crisis and its consequences on industrial production have broader economic implications within the EU ETS market. Reduced productivity and operational efficiency in participating industries not only affect their financial performance but also create a domino effect on the overall economy, potentially leading to stagnant economic growth and posing challenges for policymakers and stakeholders involved in environmental management and economic development. Additionally, the interplay between energy costs, industrial production, and economic growth influences the demand for emission permits in the EU ETS market. Decreased productivity in industries results in reduced demand for emission permits, leading to a decrease in the overall supply of available permits. This reduction in supply relative to demand exerts upward pressure on permit prices, driving them higher in the market. Understanding these interconnected factors allows stakeholders, policymakers, and market participants to navigate the complexities of the EU ETS market more effectively and make informed decisions regarding emission reduction strategies, market participation, and policy interventions aimed at achieving environmental objectives while ensuring economic sustainability. Moreover, forecasting and continuous monitoring of EU ETS prices empower companies within the system to anticipate future compliance costs, make informed investment decisions in emissions reduction technologies, and skilfully manage their carbon risk exposure. Indeed, analysts and regulators responsible for designing, implementing, and evaluating carbon pricing policies, including the EU ETS, rely on price forecasts to assess policy effectiveness, anticipate market dynamics, and make informed policy decisions. Accurate price forecasts inform the setting of emission caps, the allocation of allowances, and the adjustment of policy parameters to achieve emissions reduction targets efficiently. Finally, financial institutions in carbon trading and low-carbon projects utilise price forecasts for credit risk assessment, hedging strategies, and customised financial products. Accurate forecasts aid in offering effective risk management solutions and promoting sustainable finance initiatives.
%
%   Subsection 1.3: Literature review ----------------------------------------------   
    \subsection{Literature review}
        From an energy transition perspective, the literature offers a wide range of methods and models, providing interpretative keys and perspectives to help navigating through the challenges and opportunities described in the previous section.
        
        In their analysis, \citep{benz2009modeling} delved into the balance of supply and demand within the carbon market, categorised policy and regulation as pivotal components shaping the supply dynamics, highlighting that the carbon price is directly determined by the demand and supply of carbon allowances. To this end, subsequent studies have concluded that the price of fuel (mainly oil, natural gas and coal) is one of the most important determinants of permit prices, e.g. \citep{mansanet2007co2}, \citep{alberola2008price}, \citep{keppler2010causalities}, \citep{hintermann2010allowance}, \citep{chevallier2011model},
        \citep{creti2012carbon}, \citep{byun2013}. In these studies, economic activity is an important driver of EUA price, generally using stock market indices as indicators of economic activity. 
        In addition, also in \citep{trabelsi2023extreme}, natural gas is a key factor for EUA price. In particular, the Authors noted a dependency between the crude oil sector and the European emissions market. Furthermore, an increase in electricity prices is associated with an increase in allowance prices.
        
        In \citep{aatola2013price} the determinants of the EUA permit price are studied through a time series analysis. Companies, by producing goods and reducing emissions, influence the price of permits. Empirical data from 2005 to 2010 show a strong relationship between the price of permits and fundamentals such as electricity, gas and coal prices, confirming the importance of these variables in determining the price of EUAs. For example, \citep{hammoudeh2014energy} used a quantile regression approach to examine how fluctuations in oil, natural gas, coal and electricity prices affect 
        the distribution of EUA price. The main conclusions are that an increase in crude oil prices causes a significant drop in EUA price. Gas prices negatively affect the EUA at low levels, but positively when those are high. Electricity impacts positively at the high end of the distribution, while coal prices have a negative influence on EUA price. 
        
        Also, in \citep{hammoudeh2014explain} a Bayesian Structural Vector Autoregressive Model (BSVAR) was used to analyse the short-term dynamics of CO2 emission prices in response to changes in oil, coal, natural gas and electricity prices. The results indicate that a positive shock in crude oil prices initially increases CO2 allowance prices, but subsequently shows a negative impact; an unexpected increase in natural gas prices reduces CO2 prices; a positive shock in coal prices, the primary fuel source, has minimal and statistically insignificant effects on CO2 prices; a significant positive impact of coal prices on CO2 allowances emerges when excluding electricity prices from the BSVAR framework; and a positive shock in electricity prices negatively affects CO2 allowance prices. Furthermore, the study identifies persistent impacts of energy price shocks on CO2 allowance prices, with the most significant effects occurring six months after the shock. This effect is particularly evident in the case of shocks to natural gas and crude oil prices. In contrast, the following Authors emphasised the time-varying relationship between carbon prices and other variables. Moreover, also through a VAR model, \citep{wang2018dynamic} and \citep{ji2018information} highlight the crucial role of Brent oil yields on moving window price returns and volatility of carbon and energy prices. In particular, \citep{chevallier2009carbon} and \citep{ren2022b} explain the correlation present between permit and bond prices. \citep{tan2020connected} applied a variance decomposition to calculate the directional connection in the Carbon-Energy-Finance system, showing that the carbon market is correlated with the stock and non-energy commodity markets. 
        
        In \citep{chevallier2011model} the dynamic relationship between oil, gas and carbon prices is assessed, finding that carbon prices in Europe display a weak negative association with both oil and natural gas prices, both in Phases 2 and 3. Moreover, in \citep{chevallier2019conditional} the dependency structure between EUA yields and primary energy price yields (coal, gas, oil and electricity), modeled via a Vine copula, shows that EUA was only correlated with energy prices and that the link with oil and gas prices is negative. Furthermore, the Authors turn to the approach of Granger causality, not only to understand the relationship between  stock market and EUA spot prices, but also to achieve better forecast predictions for the future EUA values \citep{jimenez2019happens}, highlighting causality between common factors. It is observed that the causal relationship between stock markets and EUA spot prices provides valuable insights for decision-makers. 
        
        In their study, \citep{ji2018information} and \citep{wang2018dynamic} investigate an asymmetric volatility spillover effect between the EUA carbon market and the prices of WTI oil, Brent oil and EU natural gas. Focusing on how information connections and overflow effects operate between the carbon and energy markets, and analysing the interaction between returns and volatility within the carbon-energy system. Crude oil, clean energy and coal are identified as key players shaping both return and volatility patterns. In particular, the electricity market is highlighted as the main recipient of net carbon market influenced information. Additionally, overflow effects are significantly more pronounced in the volatility system than in the returns system \citep{yuan2020asymmetric}. In Phase 2 there was a strong coal spillover effect on the carbon market, while in Phase 3 natural gas became increasingly important \citep{gong2021analyzing}. 

        From a forecasting perspective, \citep{wang2021impact} apply a Bayesian network to select the most informative variables for predicting permit prices, pointing out that natural gas and crude oil directly affect the carbon price, while the S\&P500 and the Global Clean Energy Index have an indirect impact. \citep{zhao2018usefulness}, \citep{adekoya2021predicting} studied the predictive power of crude oil, natural gas and coal prices in predicting the European carbon price. The Authors found that changes in carbon prices are only weekly correlated with changes in coal prices, but are strongly correlated with natural gas.

        The problem of mixed-frequency data is common in econometrics and time series analysis to describe a situation where data is collected at different time intervals. For example, in the field of economic and financial research, it is common to have data acquired on a daily, weekly, monthly, or quarterly basis. Integrating these different data frequencies presents challenges due to the conventional assumption in time series models of a uniform frequency \citep{roberts2013gaussian}.
        
        MIDAS models \citep{ghysels2020mixed} are the most relevant parametric model used in scientific literature for combining high-frequency and low-frequency data. Mainly used to analyse and forecast macroeconomic indicators and study monetary policy effects \citep{ghysels2016macroeconomics}. MIDAS models specify a relationship between variables and estimate parameters, often assuming a specific functional form like a weighted sum or regression with lagged terms to handle mixed-frequency data. The parameters of the model are estimated using statistical techniques such as maximum likelihood estimation or Bayesian methods.
        
        To the best of our knowledge, the only effective methodology for the identification of inter-temporal information appears to be Wavelet decomposition. This technique is not widely applied in the study of financial or energy markets of a parametric type. Wavelet analysis is a mathematical technique used to examine signals and time series in both the time and frequency domains. Key parameters in Wavelet analysis include scale and translation, representing the width and position of the Wavelet, respectively. The resulting Wavelet coefficients indicate the contribution of different frequency components at various scales, facilitating the identification of significant features in the signal \citep{soltani2002use, wang2018multilevel}. In their work, \citep{spelta2023does} explore the connections between the market performance of the renewable energy sector and the fossil fuel energy sector in Europe. The study employs a multi-resolution analysis of the series using tools derived from Wavelet analysis, which breaks down time series into their time scale components associated with specific frequency ranges, proves valuable in examining the co-movements of fossil fuel and renewable index prices across various time horizons.
%
%   Subsection 1.4: Goals ----------------------------------------------------------
    \subsection{Goals}
        In an environment characterised by considerable instability it is crucial for market participants to understand the factors affecting the carbon price in order to manage market risk more effectively. Furthermore, determining whether the carbon price is influenced by fundamental or by financial nature is crucial. There is a further actor interested in carbon price analysis: the European policy maker. As already stated, the EU ETS market represents a key instrument for European climate policy, and the carbon price is a crucial indicator for assessing the effectiveness of European climate policy. It makes it possible to assess whether European policy is achieving the targets set in the Kyoto Protocol and the 2030 Agenda, and whether it needs any corrections. 

        In the context of this study, the main objective is to examine the behaviour of the carbon price and identify the factors that influence it, in order to improve market risk management by investors and to understand the substantial levers that the European policy maker can use to control the carbon price, thus making the EUA an effective tool in the fight against climate change. Motivated by previous EUA investigations, our study aims to propose new results, both empirical and methodological.

        On the empirical side, our first objective is to propose a non-parametric approach based on information theory, and specifically on the recently introduced Information Imbalance measure \citep{glielmo2022ranking}, to identify the main exogenous variables driving the EUA price. To the best of our knowledge, a similar  non-parametric analysis has never been proposed in the literature. We use Information Imbalance to also investigate the differences between the informative variables in the two Phases taken into consideration, thus verifying whether there is a disparity in price determinants between Phase 3, where the price of permits is much more stable, compared to Phase 4 where health and energy crises bring high instability to the price.

        On the methodological side, this work proposes the use of the Information Imbalance in combination with Gaussian Process regression to combine, at the most informative time scale, mixed frequency data to build forecasting or nowcasting models. Finally, our work also shows how the Information Imbalance can be used to select a small set of highly informative variables for such prediction models.
%
%   Subsection 1.5: Organization of the work ---------------------------------------
    \subsection{Organization of the work}
        The remainder of this work is organised as follows. Section \ref{sec:Data} describes the dataset used. Section \ref{sec:Methodology} introduces the essential theoretical background on Information Imbalance and Gaussian Process regression, and outlines how these are leveraged for our aims. Section \ref{sec:EmpiricalAnalysis} is centered on the empirical results, and describes the application of the Information Imbalance for the analysis of the EUA price determinants, while  \ref{sec:MethodoogyResults} focuses on the methodological results and illustrates how Information Imbalance and Gaussian Process regression can be combined to build efficient nowcasting and forecasting models on mixed-frequency data. Concluding remarks are given in Section \ref{sec:Conclusions}.
%
% Section 2: Data -------------------------------------------------------------------
\section{Data}
\label{sec:Data}
    \begin{table*}[!ht]
    \centering
    \resizebox{0.8\textwidth}{!}{%
    \begin{tabular}{ccccc}
    \hline
        \textbf{ID} & \textbf{Category} & \textbf{Variables} & \textbf{Start to end} & \textbf{Database} \\
        \hline
        0 & T & EUA (EUA) & January 2014 - April 2023 & Bloomberg\textsuperscript{\textregistered} \\ 
        \hline
        1 & UNC & GPR & January 2014 - April 2023 & GPR website \\
        2 & UNC & VSTOXX (V2X) & January 2014 - April 2023 & Bloomberg\textsuperscript{\textregistered} \\
        3 & UNC & Uncertainty EUR/USD (CAFZUUEU) & January 2014 - April 2023 & Bloomberg\textsuperscript{\textregistered} \\
        4 & UNC & Uncertainty EUR/JPY (CAFZUEJP) & January 2014 - April 2023 & Bloomberg\textsuperscript{\textregistered} \\
        5 & UNC & Uncertainty EUR/GBP (CAFZUEGB) &January 2014 - April 2023 & Bloomberg\textsuperscript{\textregistered} \\
        6 & UNC & Uncertainty EUR/CHF (CAFZUECH) &January 2014 - April 2023 & Bloomberg\textsuperscript{\textregistered} \\
        \hline
        7 & COM & ICE Dutch TTF Natural Gas (TTF0NXHR) & January 2014 - April 2023 & Bloomberg\textsuperscript{\textregistered} \\
        8 & COM & Electricity Prices Spain (OMLPDAHD) & January 2014 - April 2023 & Bloomberg\textsuperscript{\textregistered} \\
        9 & COM & Electricity Prices Germany (EXAPBDHD) & January 2014 - April 2023 & Bloomberg\textsuperscript{\textregistered} \\
        10 & COM & Electricity Prices Italy (ELIODAHD) & January 2014 - April 2023 & Bloomberg\textsuperscript{\textregistered} \\
        11 & COM & Electricity Prices France (PWNXFRAV) & January 2014 - April 2023 & Bloomberg\textsuperscript{\textregistered} \\
        12 & COM & ICE Brent oil futures (CO1 Comdty) & January 2014 - April 2023 & Bloomberg\textsuperscript{\textregistered} \\
        13 & COM & ICE Coal Rotterdam futures (TMA Comdty) & January 2014 - April 2023 & Bloomberg\textsuperscript{\textregistered} \\
        14 & COM & Gold (GCZ3 Comdty) & January 2014 - April 2023 & Bloomberg\textsuperscript{\textregistered} \\
        \hline
        15 & ER & EUR/USD spot (EUR/USD) & January 2014 - April 2023 & Eikon Refinitiv\textsuperscript{\textregistered} \\
        16 & ER & EUR/JPY spot (EUR/JPY) & January 2014 - April 2023 & Eikon Refinitiv\textsuperscript{\textregistered} \\
        17 & ER & EUR/GBP spot (EUR/GBP) & January 2014 - April 2023 & Eikon Refinitiv\textsuperscript{\textregistered} \\
        18 & ER & EUR/CHF spot (EUR/CHF) & January 2014 - April 2023 & Eikon Refinitiv\textsuperscript{\textregistered} \\
        \hline
        19 & ENR & Bloomberg Energy price return index (EUNRJP) & January 2014 - April 2023 & Bloomberg\textsuperscript{\textregistered} \\
        20 & ENR & Solactive ESG Fossil Eurozone 50 index  (S0ESG50N) & January 2014 - April 2023 & Bloomberg\textsuperscript{\textregistered} \\
        21 & ENR & S\&P Eurozone 50 Environmental index (SPEENDET) & January 2014 - April 2023 & Bloomberg\textsuperscript{\textregistered} \\
        22 & ENR & MSCI Europe Energy Sector index (MXEU0EN) & January 2014 - April 2023 & Bloomberg\textsuperscript{\textregistered} \\
        23 & ENR & ERIX index & January 2014 - April 2023 & Bloomberg\textsuperscript{\textregistered} \\
        24 & ENR & EUROSTOXX Electricity index (SXEELC) & January 2014 - April 2023 & Bloomberg\textsuperscript{\textregistered} \\
        \hline
        25 & CTRY & EUROnext100 (N100) & January 2014 - April 2023 & Bloomberg\textsuperscript{\textregistered} \\
        26 & CTRY & IBEX35 (IBEX) & January 2014 - April 2023 & Eikon Refinitiv\textsuperscript{\textregistered} \\
        27 & CTRY & DAX & January 2014 - April 2023 & Eikon Refinitiv\textsuperscript{\textregistered} \\
        28 & CTRY & CAC & January 2014 - April 2023 & Eikon Refinitiv\textsuperscript{\textregistered} \\
        29 & CTRY & FTSE Mib & January 2014 - April 2023 & Eikon Refinitiv\textsuperscript{\textregistered} \\
        \hline
        30 & MACRO & Euro-area 3-month bond yield & January 2014 - April 2023 & Bloomberg\textsuperscript{\textregistered} \\
        31 & MACRO & Euro-area 10-year bond yield & January 2014 - April 2023 & Bloomberg\textsuperscript{\textregistered} \\
        32 & MACRO & Euro-area inflation (HICP) & January 2014 - April 2023 & Eurostat \\
        33 & MACRO & Euro-area GDP (current value) & January 2014 - April 2023 & Eurostat \\
        \hline
        \multicolumn{4}{c}{\textbf{Category T}: Target; \textbf{Category UNC}: Uncertainty variables; \textbf{Category COM}: Commodity related variables; \textbf{Category ER}: Exchange rates;
        } \\
        \multicolumn{4}{c}{\textbf{Category ENR}: Energy-related indexes/variables; \textbf{Category CTRY}: Country indexes; \textbf{Category MACRO}: Macro-economic variables.}
        \\
        \hline
        \end{tabular}
        }
    \caption{\textbf{Dataset description.} List of the 34 time series used in this study, along with their source, divided in 7 categories}
    \label{Table1:VariableDescription}
    \end{table*}

    We collect daily closing prices of EUA from Bloomberg\textsuperscript{\textregistered}, spanning from January 2014 to April 2023. This dataset consists of 2374 observations, extending beyond the period examined in \citep{wang2023}. Consistent with previous studies on EUA price, we exclude both Phase 1 and Phase 2 of the market from our analysis. This decision is justified by the evidence that during Phase 1 and Phase 2, price fluctuations are known to have been primarily influenced by regulatory and policy changes, given that the market was still in a testing phase.

    Following other existing studies such as \citep{chevallier2011macroeconomics}, \citep{byun2013}, \citep{tan2021}, \citep{ren2022a, ren2022b}, \citep{aller2021robust}, we categorise the 33 predictors into 6 categories, relating to geopolitical, economic, and financial uncertainty, commodities, some exchange rates, energy indices, national indices, the main European index, and finally, macroeconomic variables. All predictors were collected daily, except for Euro-area Inflation collected monthly and Euro-area GDP collected quarterly. The reference period is the same as the one of the target variable, with the same number of observations for daily-collected predictors and with 114 and 37 observations for monthly Inflation and quarterly GDP, respectively. First, we consider 6 predictors related to uncertainty, namely: (1) GeoPolitical Risk (GPR) index; (2 - 6) uncertainty indexes of major world exchange rates: EUR/USD, EUR/JPY, EUR/GBP, EUR/CHF. The uncertainty factors are collected on Bloomberg, while the GPR is collected from its dedicated website \citep{gpr}. Second, we include the following 8 commodities not necessarily related to energy: (7) ICE Dutch natural gas futures; (8 - 11) electricity prices for Spain, Germany, Italy, and France; (12) ICE Brent oil futures; (13) ICE Rotterdam coal futures, and finally (14) the gold index. Third, we consider major spot interest rates to explore the possible purely financial and highly volatile effects, (15 - 18): EUR/USD, EUR/JPY, EUR/GBP, EUR/CHF. Fourth, we include the following 6 European energy indices to explore the informational content of energy indices of different nature: (19) Bloomberg Energy price return index; (20) Solactive ESG Fossil Eurozone 50 index; (21) S\&P Eurozone 50 Environmental index; (22) MSCI Europe Energy Sector index; (23) ERIX index; (24) EUROSTOXX Electricity index. Then, we consider the influence of 4 industrial countries and one European index to explore the predictive content of financial activities: (25) EUROnext100; (26) IBEX35; (27) DAX; (28) CAC; (29) FTSE Mib. Finally, we consider the influence of the economic cycle and European macroeconomic conditions, considering: (30) Euro-area 3-month bond yield; (31) Euro-area 10-year bond yield; (32) Euro-area inflation; (33) Euro-area GDP. 

    As already noted, variables (32) and (33) are collected at a  frequency different from the daily one of the EUA target variable, and hence require a degree of data imputation to facilitate the examination of their impact. These imputations can incur in errors and lead to information loss. For this reason we decided to keep these two variables out of our empirical analysis and to use only variables that have undergone no transformation. However, Euro-area inflation and Euro-area GDP are included in the analysis we perform in Section \ref{sec:MethodoogyResults}, where we put forward a non-parametric approach to overcome this mixed frequency problem. Our variable selection process involved a comprehensive review of literature and deep market understanding. By introducing new metrics such as GDP and Inflation alongside traditional ones, we successfully address the challenge of mixed frequency in studying the macroeconomic dynamics with EUA. Furthermore, considering macroeconomic indicators like GDP and inflation when studying EU ETS price determinants is essential for gaining insight into the broader economic landscape, discerning market sentiments, and anticipating policy implications for emissions trading \citep{konradt2024carbon}.
%
% Section 3: Methods ----------------------------------------------------------------
\section{Methods}\label{sec:Methodology}
%   Subsection 3.1: The Information Imbalance ---------------------------------------
    \subsection{The Information Imbalance}
        The Information Imbalance is a measure recently introduced to quantify the degree to which one or more variables can be used for predicting another set of variables \citep{glielmo2022ranking}. The Information Imbalance can be formulated in information-theoretic terms using the theory of copulas as explained in \citep{glielmo2022ranking}. For brevity, we will not review its theoretical underpinning but rather its practical definition and interpretation.

        Given a variable $X$ and any two points (or statistical units) $i$ and $j$, the \emph{rank} $r^{X}_{ij}$ of $j$ with respect to $i$ is obtained by sorting pairwise distances between $i$ and all other points in ascending order. The \emph{rank} $r^{X}_{ij}$ is the position of the distance between $i$ and $j$ in the ordered sequence.
        For example, $r^{X}_{ij}=1$ if point $j$ is the closest to point $i$ according to the distance $d_{X}$. Similarly,  $r^{Y}_{ij}$ is the rank of $j$ with respect to $i$ according to distance $d_Y$ and, in general $r^{Y}_{ij} \neq r^{X}_{ij}$. The Information Imbalance \emph{from $X$ to $Y$}, $\Delta(X \rightarrow Y)$, can then be defined on a dataset with $N$ points on which we record joint values for both $X$ and $Y$, as 
        \begin{equation}
        \label{eq:imbalance}
            \Delta (X\rightarrow Y) = \frac{2}{N} \mathbb{E}\left[r_Y\mid r_X=1\right].
        \end{equation}
        where the expected value is taken only over the nearest neighbour points according to variable $X$. By construction, in the limit of $N \rightarrow \infty$, the Information Imbalance is statistically confined in the interval $(0,1)$,
        with $\Delta (X\rightarrow Y)\approx0$ implying that $X$ is fully informative for $Y$ and, conversely, $\Delta (X\rightarrow Y)\approx1$ implying that $X$ carries no information useful for predicting $Y$. This limiting behaviour is readily understood from Eq. \ref{eq:imbalance}. For identical variables, the expected value in the equation evaluates to $1$, and the Imbalance evaluates to $2/N$, a number close to zero for large enough $N$. For completely unrelated variables, the expected value evaluates to $N/2$ and the Imbalance to $1$, on average.

        \begin{figure}[htp]
        \centering
            \includegraphics[width=0.99\columnwidth]{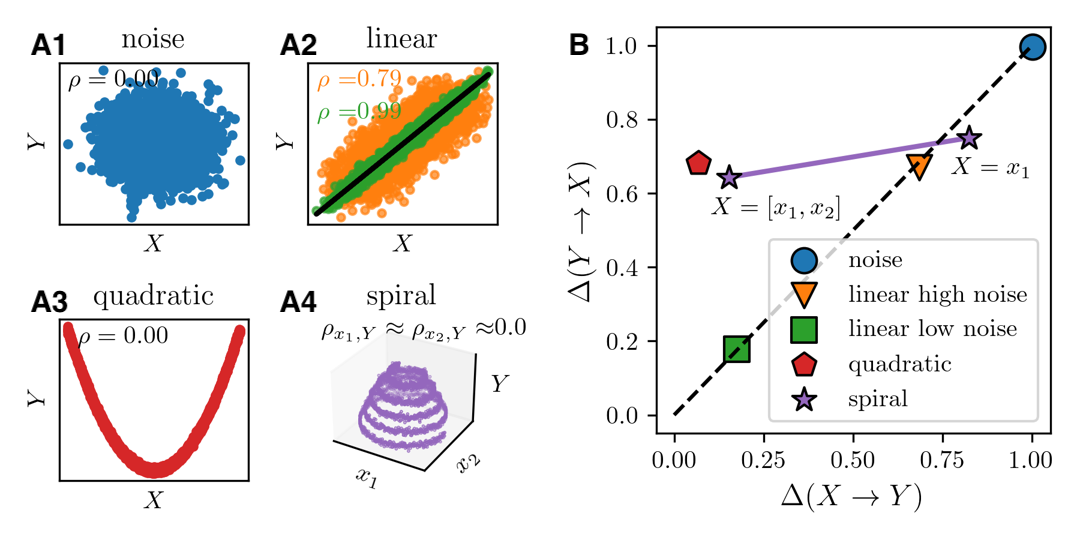}
            \caption{\textbf{The Information Imbalance.} An illustration of the Information Imbalance computed for a number of intuitive relationships between independent variables $X$ and a dependent (target) variable $Y$. Specific datasets in the \textbf{A} panels correspond to markers of the same colour in the \textbf{B} panel. Note that the Information Imbalance well captures the trivial relationships of \textbf{A1} and \textbf{A2}, as well as the nonlinear (quadratic) relationship of \textbf{A3} and the multivariate relationship existing in \textbf{A4}, for which the linear correlation coefficient $\rho$ cannot be used.}
        \label{fig:TheInformationImbalance}
        \end{figure}

        The Information Imbalance plane is a plot of $\Delta (X\rightarrow Y)$ vs $\Delta (Y \rightarrow X)$. A point in such a plane represents the relationships between any two variables  $(X, Y)$. This is illustrated in Figure \ref{fig:TheInformationImbalance}, where 4 types of synthetic datasets are analysed through the Information Imbalance plane. Specifically, in panel \textbf{A1} variables $X$ and $Y$ are related purely by Gaussian noise, their Pearson correlation $\rho$ is zero, and the $\Delta(X \rightarrow Y) = \Delta (X \rightarrow Y) = 1$, 
        resulting in a blue circle in the top right of the Information Imbalance plane.
        In \textbf{A2} some correlation exists between the two variables, and this results in a higher value of the Pearson correlation $\rho$ and in a gradual shift in the Information Imbalance plane from one towards zero along the diagonal of the plane. 
        The lower is the noise level around the linear relation between $X$ and $Y$ the closer to zero is the corresponding point in the Information Imbalance plane: compare the orange triangle with the green square.
        These linear relationships are trivially captured by the Pearson correlation coefficient $\rho$ as well as by the Information Imbalance which, however, can also be used to probe much more complex dependencies (both nonlinear and multivariate) as illustrated in \textbf{A3} and \textbf{A4}. In \textbf{A3} , the nonlinear (quadratic) relationship results in a linear correlation of zero, but gives rise to a low value of the Information Imbalance $\Delta(X \rightarrow Y)$ and a high value of $\Delta(Y \rightarrow X)$ indicating that $Y$ can be predicted by $X$ better than the opposite. In \textbf{A4}, a multivariate relationship exists between $x_1$, $x_2$ and $Y$ whereby two variable combination [$x_1$, $x_2$] is much more useful in predicting $y$ than any one of the two taken singularly. Once again the linear correlation $\rho$ cannot capture such an effect, while the Information imbalance $\Delta(X \rightarrow Y)$ drastically decreases as $x_2$ is added to $x_1$ in the set of explanatory variables: compare the two purple stars in Figure \ref{fig:TheInformationImbalance}.
%
%   Subsection 3.2: Gaussian Process regression -------------------------------------
    \subsection{Gaussian Process regression}
    \label{ssec:GPs}
        A Gaussian Process (GP) is a powerful and versatile statistical tool used in various fields, including machine learning \citep{williams2006gp}, statistics \citep{shi2020gp} or Bayesian optimisation \citep{wilson2016gp}. GPs have gained popularity due to their flexibility and their effectiveness in quantifying uncertainty, they are a well-known approach to representing functions in a non-parametric setting along with neural networks.

        A GP can be conceived as an infinite-dimensional generalisation of a multivariate Gaussian distribution. More precisely, a GP is a collection of random variables, where any finite subset of them follows a multivariate Gaussian distribution. To define a GP we first need to select a mean function $\mu(\mathbf{x})$, which provides the expected value of the modelled function $\mu(\mathbf{x})=\mathbb{E}\left[f(\mathbf{x})\right]$. Without loss of generality, the GP mean function is typically assumed to be zero as the GP is typically applied to standardised data. We adopt this common practice throughout this work. The second and fundamental component of a GP is the covariance function (or kernel), a function that characterises the relationships between different points in the function's domain and quantifies the correlation or similarity between function values at different input points:
        \begin{equation} \label{Equation_2}
            k \left( \mathbf{x}, \mathbf{x}' \right) =\mathbb{E }\left[f(\mathbf{x}) f(\mathbf{x}') \right].
        \end{equation}
    The choice of the GP's kernel function should take into account the relationships or dependencies between data from different sources. We tested various kernel functions for this work, such as Radial Basis Function (RBF), additive, multiplicative with and without constants, and Matern kernel, ultimately selecting the latter, defined as
        \begin{equation}
        \label{ed:matern_kernel}
            k_{Matern}(\mathbf{x}, \mathbf{x}') = \frac{1}{\Gamma(\nu)2^{1-\nu}}\left(\sqrt{2\nu}\frac{\|\mathbf{x}-\mathbf{x}'\|}{l}\right)^\nu K_\nu\left(\sqrt{2\nu}\frac{\|\mathbf{x}-\mathbf{x}'\|}{l}\right).
        \end{equation}
        In the Matern kernel, the $\nu$ parameter controls the smoothness, and the $l$ parameter controls the length scale of variations of the resulting function. Once a covariance function is selected, the GP can be used as a prior distribution and can be fit to a dataset $\mathcal{D} = \{ (\mathbf{x}_i, y_i) \}_{i=1}^N$ in Bayesian regression called Gaussian Process regression. The posterior distribution in a Gaussian Process regression is also a GP and the posterior mean, the curve that best fits the data, can be computed analytically
        \begin{equation}
        \label{eq:gp_predictive_mean}
            \mu(\mathbf{x}^*) = \mathbf{k}_*^T (\mathbf{K} + \sigma_n^2 \mathbf{I})^{-1} \mathbf{y},
        \end{equation}
        where $(\mathbf{K})_{ij} = k(\mathbf{x}_i, \mathbf{x}_j)$ is an entry in the kernel matrix, $(\mathbf{k}_*)_i = k(\mathbf{x}_i, \mathbf{x}^*)$ is the kernel between the dataset and the test point, and $\sigma_i^2$ controls the level of noise that is assumed to be present in the data.
        \begin{figure}[htp]
        \centering
            \includegraphics[width=0.9\columnwidth]{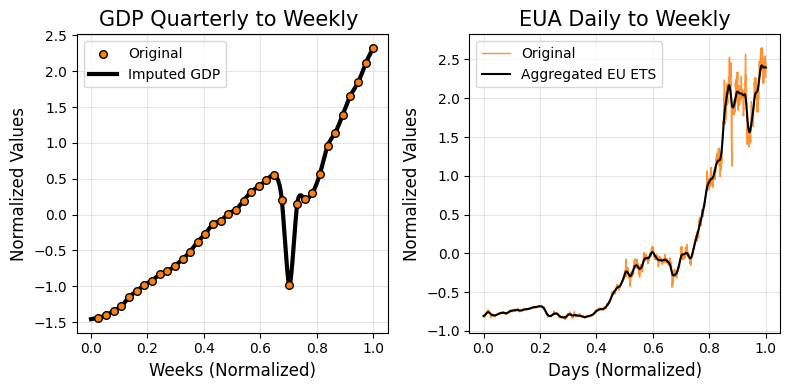}
            \caption{\textbf{Imputation and aggregation using GPs.} In the left panel, a GP is used to impute the GDP time series variable from quarterly to weekly frequency. In the right panel, a GP is used to aggregate the target time series of EUA price from daily to weekly frequency.}
        \label{fig:AggregationImputation}
        \end{figure}
%
%   Subsection 3.3: Information Imbalance and Gaussian Processes to analyse EUA price
    \subsection{Information Imbalance and Gaussian Processes for the\\ analysis EUA price}
    \paragraph{\textbf{Price determinants}}
        The Information Imbalance represents a rather natural tool to use to answer the question of what determines the price of the EUA. In this work we use the Imbalance $\Delta(X_{t}\rightarrow\text{EUA}_{t+\delta t})$ to quantify the information that a predictor set $X_t$ at time $t$, which can encompass any combination of variables in Table \ref{Table1:VariableDescription}, contains on the target variable $\text{EUA}_{t+\delta t}$, at time ${t+\delta t}$. We use the implementation of the Information Imbalance available in the DADApy package \citep{dadapy}. In the quest for informative variables, Wavelet decomposition stands as a parametric alternative to Information Imbalance. However, it comes with drawbacks. Firstly, it significantly inflates the dimensionality of the feature space, posing computational hurdles and necessitating additional steps for dimensionality reduction \citep{liu2009wavelet}. Moreover, its computational demands are magnified, particularly with high-dimensional data, impacting the efficacy of feature selection algorithms \citep{pati1993orthogonal}. On the other hand, Information Imbalance, rooted in a non-parametric framework, circumvents dimensionality inflation, maintaining lower computational intensity. It furnishes robust outcomes even amidst the challenges of outliers and noise.
    \paragraph{\textbf{Mixed frequency forecasting}}
        We propose a non-parametric method that leverages Gaussian Processes in conjunction with the Information Imbalance to optionally aggregate data from different time frequencies. An example of imputation and aggregation of time series using GPs is shown in Figure \ref{fig:AggregationImputation}. For the imputation --left panel of the figure-- a GP is fit to the data using a low noise level of $\sigma_n^2=10^{-3}$ just needed to regularise the inversion in Eq. \ref{eq:gp_predictive_mean}, and the posterior GP mean is then used to compute the value of the time series at the needed frequencies. For the aggregation --right panel of the figure-- the GP noise level is set to the average rolling variance of the time series computed using the target period as the rolling window; and the posterior GP mean is similarly used to compute the value of the time series at the needed frequencies. We also used GPs to perform experiments of nowcasting and forecasting of allowance prices.

        Additionally, a $k$-fold cross-validation with $k=5$ was applied. In all our experiments, we set the smoothness parameter, $\nu$, to $1.5$, the largest degree of smoothness compatible with all the time series available, and we select the length scale, $l$, separately for every fit using a maximum likelihood optimisation \citep{williams2006gp}. As noted in \citep{williams2006gp}, a GP is a stochastic process where any finite set of points follows a joint Gaussian distribution. However, this doesn't mean the data distribution itself is Gaussian. Each point in a GP corresponds to a Gaussian-distributed random variable, defined by its mean and covariance functions. These functions characterise the process but don't dictate the data's distribution. GPs are often used as priors in Bayesian inference. When combined with observed data likelihood, the resulting posterior distribution over functions may or may not be Gaussian, as in our case.
%
% Section 4: Price determinants -----------------------------------------------------
\section{Price determinants} \label{sec:EmpiricalAnalysis}
%   Subsection 4.1: Descriptive statistics and correlation analysis -----------------
    \subsection{Descriptive statistics and correlation analysis}
        \begin{table*}[ht]
        \centering
        \resizebox{0.8\textwidth}{!}{%
        \begin{tabular}{ccccccccc}
            \hline
            \textbf{ID} & \textbf{Variables} & \textbf{Mean} & \textbf{STD} & \textbf{Min} & \textbf{25\%} & \textbf{50\%} & \textbf{75\%} & \textbf{Max} \\
            \hline
            0 & EUA & 27.31 & 27.55 & 3.93 & 6.41 & 17.59 & 32.59 & 100.29 \\
            \hline
            1 & GPR & 113.50 & 52.80 & 9.49 & 79.90 & 103.89 & 136.51 & 542.66 \\
            2 & VSTOXX & 20.87 & 7.38 & 10.68 & 15.88 & 19.45 & 24.03 & 85.62 \\
            3 & Unc. EUR/USD & 2.66 & 0.60 & 1.57 & 2.21 & 2.57 & 3.05 & 4.28 \\
            4 & Unc. EUR/JPY & 2.96 & 0.60 & 1.32 & 2.64 & 2.99 & 3.31 & 5.69 \\
            5 & Unc. EUR/GBP & 2.42 & 0.64 & 1.36 & 1.94 & 2.37 & 2.72 & 6.64 \\
            6 & Unc. EUR/CHF & 1.85 & 0.73 & 0.98 & 1.49 & 1.74 & 1.98 & 8.92 \\
            \hline
            7 & Natural Gas & 33.90 & 41.47 & 3.63 & 14.90 & 19.15 & 24.57 & 311\\
            8 & Elec. Prices Spain & 69.82 & 53.82 & 1.10 & 42.61 & 52.34 & 65.55 & 544.98 \\
            9 & Elec. Prices Germany & 69.23 & 83.68 & -9.12 & 31.53 & 39 & 58.61 & 682.89 \\
            10 & Elec. Prices Italy & 92.18 & 96.61 & 10.66 & 46.95 & 55.67 & 74.10 & 718.71 \\
            11 & Elec. Prices France & 92.18 & 96.61 & 10.66 & 46.95 & 55.67 & 74.10 & 718.71 \\
            12 & Brent oil & 66.97 & 21.97 & 17.32 & 50.19 & 63.49 & 79.17 & 133.89 \\
            13 & Coal futures & 118.37 & 97.02 & 48.50 & 64.51 & 82.48 & 109.83 & 457.80 \\
            14 & Gold & 1,457.71 & 274.68 & 1,051.10 & 1,240.92 & 1,319.00 & 1,760.34 & 2,063.54 \\
            \hline
            15 & EUR/USD spot & 1.15 & 0.08 & 0.96 & 1.10 & 1.13 & 1.18 & 1.39 \\
            16 & EUR/JPY spot & 129.74 & 8.11 & 111.15 & 123.42 & 129.68 & 135.82 & 149.18 \\
            17 & EUR/GBP spot & 0.84 & 0.05 & 0.69 & 0.83 & 0.86 & 0.88 & 0.94 \\
            18 & EUR/CHF spot & 1.10 & 0.06 & 0.95 & 1.06 & 1.09 & 1.14 & 1.24 \\
            \hline
            19 & Bloomberg Energy price return index & 1.10 & 0.06 & 0.95 & 1.06 & 1.09 & 1.14 & 1.24 \\
            20 & Solactive ESG Fossil Eurozone 50 index & 101.58 & 17.01 & 48.01 & 91.17 & 106.05 & 113.87 & 130.32 \\
            21 & S\&P Eurozone 50 Environmental index & 127.29 & 26.58 & 84.21 & 107.34 & 122.45 & 139.70 & 199.47 \\
            22 & MSCI Europe Energy Sector index & 1,477.89 & 152.24 & 1,059.38 & 1,350.86 & 1,467.14 & 1,560.97 & 1,870.90 \\
            23 & ERIX index & 1,325.94 & 617.13 & 567.78 & 840.82 & 1,028.58 & 1,949.09 & 3,106.55 \\
            24 & EUROSTOXX Electricity index & 314.53 & 67.26 & 212.46 & 253.71 & 284.09 & 379.67 & 471.78\\
            \hline
            25 & EUROnext100 & 1,034.78 & 156.67 & 733.93 & 899.78 & 1,016.07 & 1,133.29 & 1,388.09 \\
            26 & IBEX35 & 1,034.78 & 156.67 & 733.93 & 899.78 & 1,016.07 & 1,133.29 & 1,388.09 \\
            27 & DAX & 12,216.79 & 1,870.47 & 8,441.71 & 10,685.23 & 12,238.56 & 13,263.39 & 16,271.75\\
            28 & CAC & 5,337.46 & 840.69 & 3,754.84 & 4,615.03 & 5,234.00 & 5,882.29 & 7,577\\
            29 & FTSE Mib & 21,623.98 & 2,699.71 & 14,894.44 & 19,727.58 & 21,674.43 & 23,329.29 & 28,162.67\\
            \hline
            30 & Euro-area 3-month bond yield & -0.38 & 0.68 & -1.13 & -0.71 & -0.59 & -0.29 & 4.55 \\
            31 & Euro-area 10-year bond yield & 0.39 & 0.74 & -0.85 & -0.21 & 0.32 & 0.68 & 2.75 \\
            \hline
        \end{tabular}
         }
        \caption{\textbf{Descriptive statistics.} Mean, standard deviation, minimum, maximum and three percentiles of the 34 time series considered in this work and also listed in Table \ref{Table1:VariableDescription}.}
        \label{Table2:DescpritiveStatistics}
        \end{table*}
        Table \ref{Table2:DescpritiveStatistics} shows the descriptive statistics of all predictors and EUA, showing the mean, the standard deviation (STD), the minimum, the 0.25, 0.5 and 0.75  percentiles and the maximum value.
    \begin{figure*}[h]
        \centering
        \includegraphics[width=0.98\textwidth]{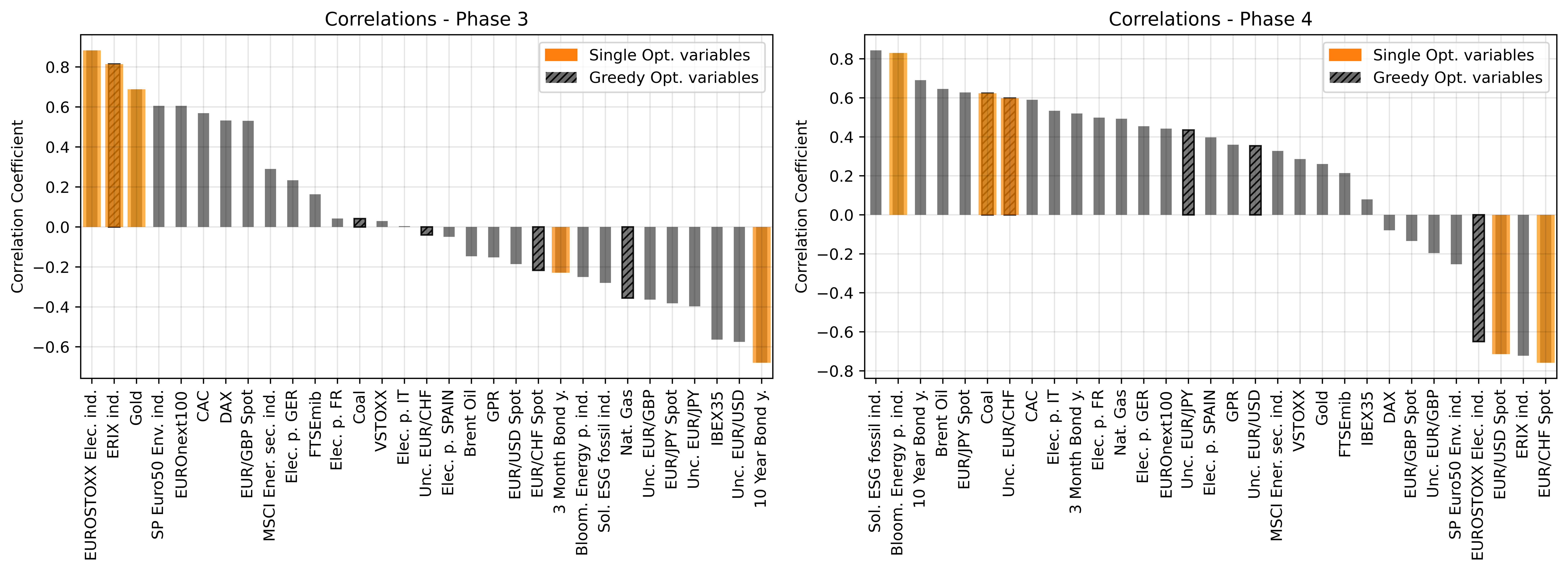}
        \caption{\textbf{Correlation analysis.}  Correlations between the raw daily data and EUA. The most informative single variables selected using the Information Imbalance (Fig. \ref{fig:SingleVarDaily}) are highlighted in orange. The dashed bars refers to the 5 most informative variables obtain through a greedy selection algorithm based on the Information Imbalance (Fig. \ref{fig:GreedyVar_Daily}), for Phase 3 (left) and Phase 4 (right).}
        \label{fig:Correlations}
    \end{figure*}
        In Figure \ref{fig:Correlations} we present the correlations between the explanatory variables and the EUA price, for each phase. We notice that the variables belonging to the commodity category turn out to be highly correlated with the EUA price; this is a predictable result as they share the same category. In contrast, during Phase 3, the uncertainty-related variables are poorly correlated with the target variable. Especially during Phase 4 most of the considered variables  have a positive correlation, with a large number of them being highly correlated. The exchange rates are not very correlated with our target variable in Phase 3, although some of these rates, as EUR/CHF Spot and EUR/USD Spot are strongly negatively correlated in Phase 4. Finally, we can observe that the correlation between EUROSTOXX Electricity prices index, ERIX index and S\&P Euro50 Environmental index, undergoes a substantial change, shifting from a strong positive correlation in Phase 3 to a strong negative correlation in Phase 4. Notably, the correlation analysis conducted in this section serves solely as an initial assessment of the variables in our dataset.
%
%   Subsection 4.2: Information Imbalance analysis ----------------------------------
    \subsection{Information Imbalance analysis}
        \begin{figure*}[h]
        \centering
            \includegraphics[width=0.98\columnwidth]{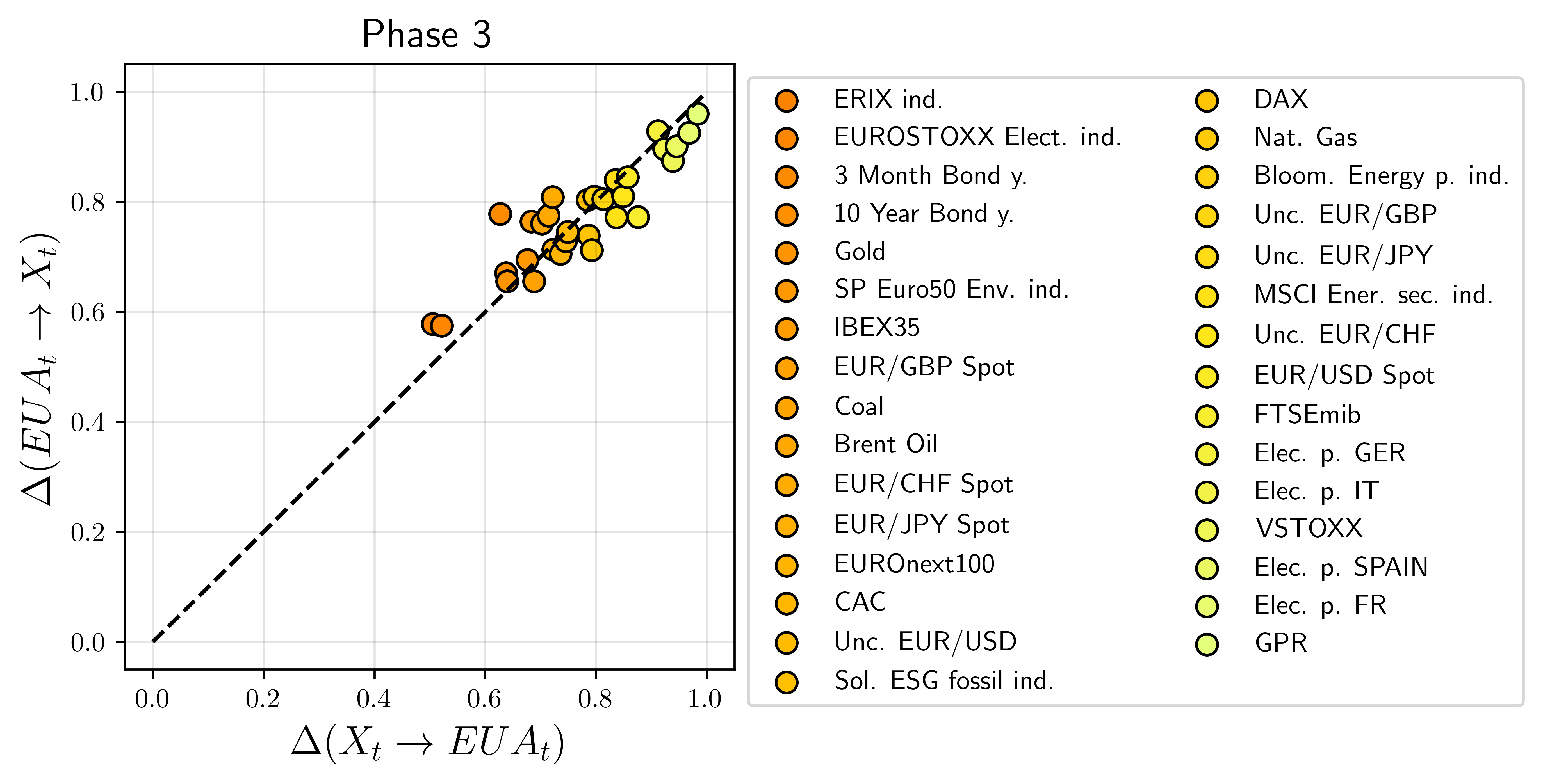}
            \includegraphics[width=0.98\columnwidth]{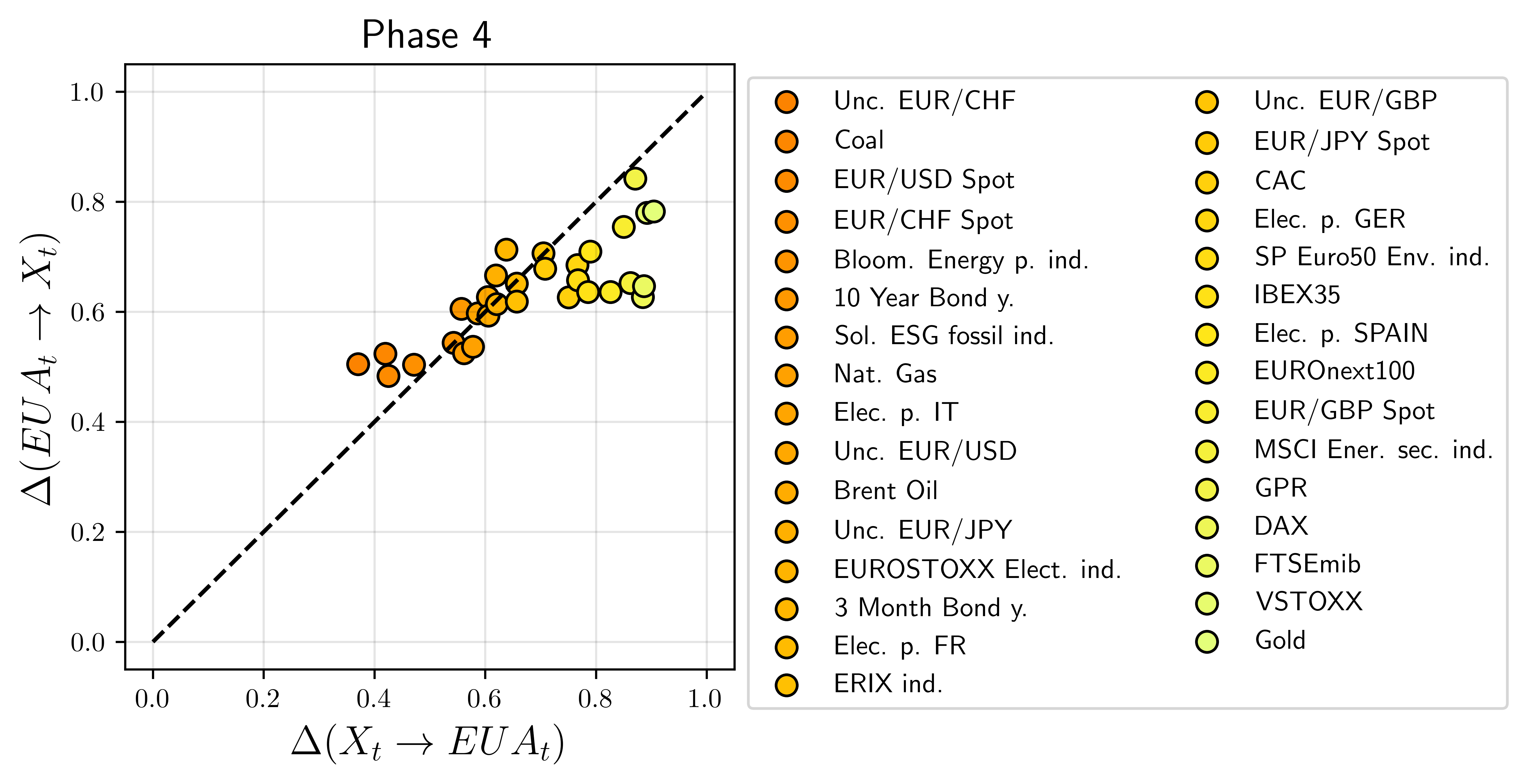}
            \caption{\textbf{Daily Information Imbalance analysis.} This plot analyses the information content of each variable in our dataset taken individually. The legend is order from the most informative (darker orange) to the less informative variables (yellow orange), for Phase 3 (left) and Phase 4 (right).}
        \label{fig:SingleVarDaily}
        \end{figure*}
        Unlike traditional parametric econometric models, the Information Imbalance does not require any assumption on the underlying data-generating process but rather allows working with the variables as they were collected so as not to alter the results in any way. This is undoubtedly one of the main advantages of the adopted non-parametric approach, which provides a lower level of restriction than any other parametric analysis. 

        In Figure \ref{fig:SingleVarDaily}, the two Information Imbalance planes for Phase 3 and Phase 4 are shown. In these two plots, the Information Imbalance between individual explanatory variables $X_t$ and the target variable $EUA_t$ are reported. On the $x$-axis we find the Information Imbalance from the predictors set towards the target set ($\Delta(X_t \rightarrow EUA_t $)), while the opposite relationship ($\Delta(EUA_t\rightarrow X_t$)) is presented on the $y$-axis. It is interesting to compare the two and observe the main differences between the two phases in terms of informative variables. 

        As far as Phase 3 is concerned, the single most informative variable is the ERIX index which monitors the progress of European renewable energy companies involved in one or more of six investment clusters, which include biofuels, geothermal, marine, solar, hydro and wind energy. \citep{kanwal2021} demonstrated how EUA shares are independent of the ERIX index by analysing their time-varying correlation using a GO-GARCH model. However, our study shows that the information content of the ERIX index relative to EUA is very high compared to the other variables taken into account, opening up possible new interpretations. In particular, we can say that the price behaviour of the ERIX index is close to that of the EUA, suggesting similar market dynamics. Although renewable energy sources do not participate directly in emissions trading, their importance lies in shaping the broader dynamics of the market \citep{chun2022relationship}. The use and promotion of renewable sources contribute to an overall reduction of greenhouse gas emissions in the energy sector \citep{hailemariam2022does}. Consequently, they complement the objectives of the EU Emissions Trading Scheme, which aims to reduce emissions from industrial activities. In some situations, renewable energy projects generate carbon credits or offsets, symbolising the reduction or avoidance of greenhouse gas emissions. Companies participating in the EU emissions trading system can use these credits to offset a portion of their emissions, thus fulfilling compliance obligations more effectively. The demand for emission allowances in the EUA is influenced by the energy composition \citep{hanif2021nonlinear}. Increased use of renewable energy sources can lead to reduced emissions from the power generation sector, thus influencing the supply and demand dynamics in the emissions trading market. The EU is actively working on integrating renewable energy policies with emission reduction targets. For example, the \citep{eudirective} sets binding targets for the share of renewable energy in the EU's final energy consumption. This integration ensures that efforts to promote renewable energy are in line with broader climate goals, including those of the EU Emissions Trading Scheme. Furthermore, the EU Emissions Trading Scheme creates economic incentives for investments in cleaner technologies, including renewable energy projects. Companies investing in renewable energy can benefit not only from the sale of clean energy, but also from potential gains from the sale of emission allowances or carbon credits \citep{euets}. In addition, in Phase 3, we see that the most informative individual variables are all intertwined with the European energy market, such as the EUROSTOXX Electricity index. We also find variables such as 3 month and 10 year bond yields which represent the interest rate required by investors to hold Eurozone government bonds with a maturity of 10 years and 3 months respectively. This indicator is significant because changes in bond yields provide valuable information on market expectations related to economic conditions, inflation and monetary policy \citep{altavilla2014bond}.
        \begin{figure}[h]
        \centering
        \includegraphics[width=0.98\columnwidth]{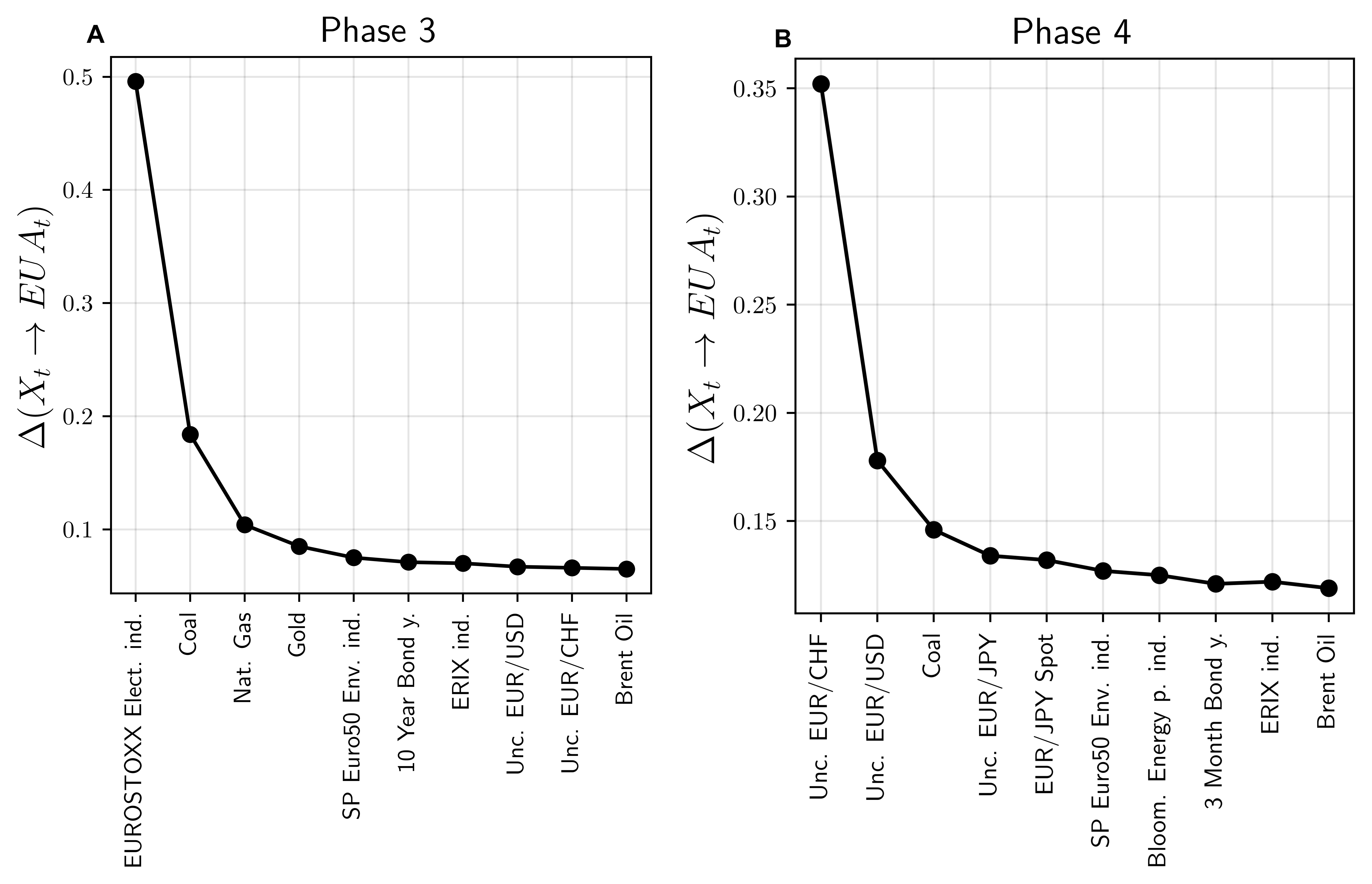}
        \includegraphics[width=0.98\columnwidth]{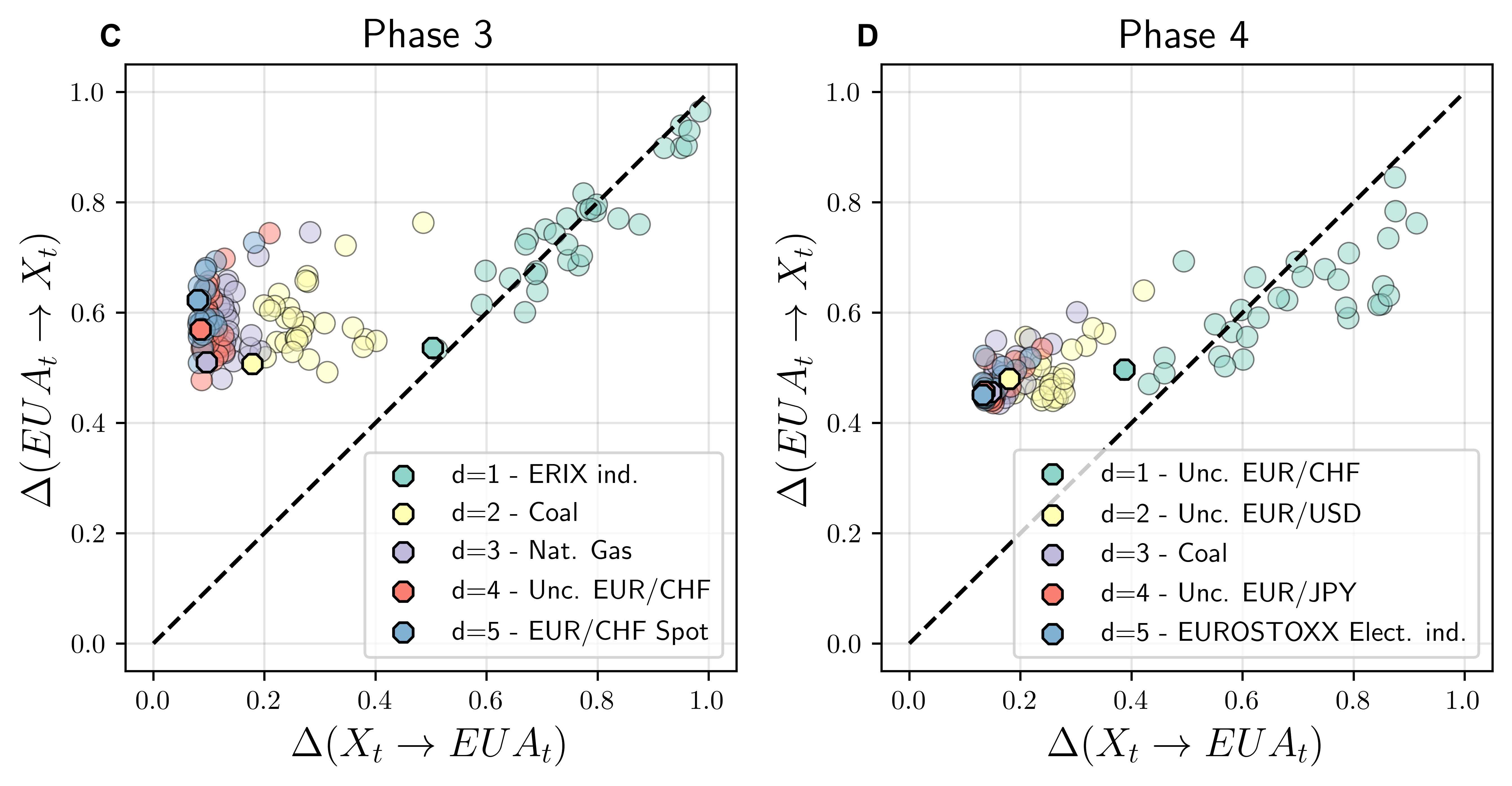}
        \caption{\textbf{Information Imbalance analysis of Phase 3 and Phase 4 EUA price determinants.} 
        The first row shows the Imbalance $\Delta(X_t \rightarrow EUA_t)$ from the predictor set to the EUA price, for a growing number of variables in $X_t$, with the labels on the x-axis indicating the variable being added.
        The second row depicts the greedy optimisation used to select the most informative variables on the Information Imbalance plane.
        }
        \label{fig:GreedyVar_Daily}
    \end{figure}
        In Phase 4, we observe a situation that differs but bears resemblances in certain aspects. In particular, we can appreciate how variables reflecting financial fluctuations turn out to have a much greater importance than in Phase 3. Uncertainty regarding an exchange rate, as in the case of the EUR/CHF uncertainty, the most informative variable at this phase, refers to the lack of predictability or confidence in the future movements of a currency pair. High volatility in the exchange rate signals heightened uncertainty.  Exchange rate options (in particular, implied volatility) offer insights into market participants' expectations on future currency movements \citep{beckmann2017exchange}. Changes in global risk sentiment, often reflected in stock market movements, can influence demand for safe-haven currencies. Uncertainty about global economic conditions can lead to increased volatility in currency markets. Extensive speculative trading or sudden changes in market sentiment can contribute to uncertainty. Rapid changes in market sentiment based on the actions of speculators can lead to unpredictable currency movements \citep{ferrara2022measuring}. 

        This difference in the selection of informative variables shows how the impact of the COVID-19 pandemic and the energy crisis completely disrupted the price dynamic, defining new determinants of the EUA price. Industries covered by the EUA may explore the possibility of transitioning from coal to cleaner energy alternatives, such as natural gas or renewables, driven by economic considerations and environmental goals. The economic viability of these substitutions can be affected by changes in coal prices \citep{bohringer2022europe}. Although both markets share similar influences, the two markets are not directly connected but can be influenced by broader dynamics in the energy market, including shifts in supply and demand, geopolitical events, and economic conditions \citep{anke2020coal}. Regulatory changes related to carbon emissions and coal use can affect both markets, with stringent regulations potentially increasing costs and affecting demand for EUA permits. The transition to renewable energy sources is a significant factor, which could reduce demand for coal and influence the carbon market. Both markets are also influenced by global factors such as international trade, energy prices and climate change policies \citep{chun2022relationship}. 

        The changing energy landscape, with an increasing focus on cleaner and more sustainable sources, has the potential to change the demand for coal and the assessment of EUA permits. Renewable energies, such as wind, solar, hydro and bio-energy, represent low-carbon or carbon-neutral energy sources. The use and promotion of these renewable sources contribute to an overall reduction of greenhouse gas emissions in the energy sector \citep{hailemariam2022does}. The demand for emission allowances in the EUA is influenced by the energy composition \citep{hanif2021nonlinear}. Increased use of renewable energy sources can lead to reduced emissions from the power generation sector, thus influencing the supply and demand dynamics in the emissions trading market. The EU is actively working on integrating renewable energy policies with emission reduction targets. For example, the \citep{eudirective} sets binding targets for the share of renewable energy in the EU's final energy consumption. It is crucial to recognise that although correlations exist, each phase is subject to unique factors.
%
%   Subsection 4.3: Greedy selection of variables -----------------------------------
    \subsection{Greedy selection of variables}
    \label{subsec:greedy_selection}
        Figure \ref{fig:GreedyVar_Daily} shows the result of an iterative greedy procedure to select the set of explanatory variables that is most informative about the EUA price. The algorithm starts with an empty set and then iteratively adds to the set the variable that brings the highest information content given the current set composition. The process continues until a sufficiently informative subset of variables is obtained. The most informative subset for Phase 3 turns out to contain the following explanatory variables: ERIX index, Coal and Natural Gas. Looking beyond these three variables is not advisable since the improvement in information content is negligible. This effect can be seed in Figure \ref{fig:GreedyVar_Daily}, where the Information Imbalance is plotted as a function of the number of variables (A, B), and from the corresponding Information Imbalance plane (C, D) by observing the higher concentration of dots as more variables are added to the set.

        For Phase 4, the most informative subset turns out to be different with the exception of Coal. The distinctive feature we notice in this instance is the presence of the uncertainty indicators computed over the exchange rates considered in this study. Uncertainty in exchange rates can contribute to market-wide uncertainties. Investors often regard currency fluctuations as a risk, and elevated uncertainty in exchange rates may prompt increased risk aversion among investors, potentially influencing their behaviour in the EU ETS market. Central banks' actions in response to uncertainties in exchange rates and changes in monetary policies can shape interest rates and broader economic conditions. These dynamics, in turn, may have implications for the regulatory framework and policy decisions related to emissions trading within the EUA \citep{chevallier2011options}.
%
% Section 5: Time-scale aggregation and forecasting -----------------------------------
\section{Time-scale aggregation and forecasting}
\label{sec:MethodoogyResults}
%   Subsection 5.1: Data frequency selection ------------------------------------------
    \subsection{Data frequency selection}
        \begin{figure}[h]
        \centering
            \includegraphics[width=0.98\columnwidth]{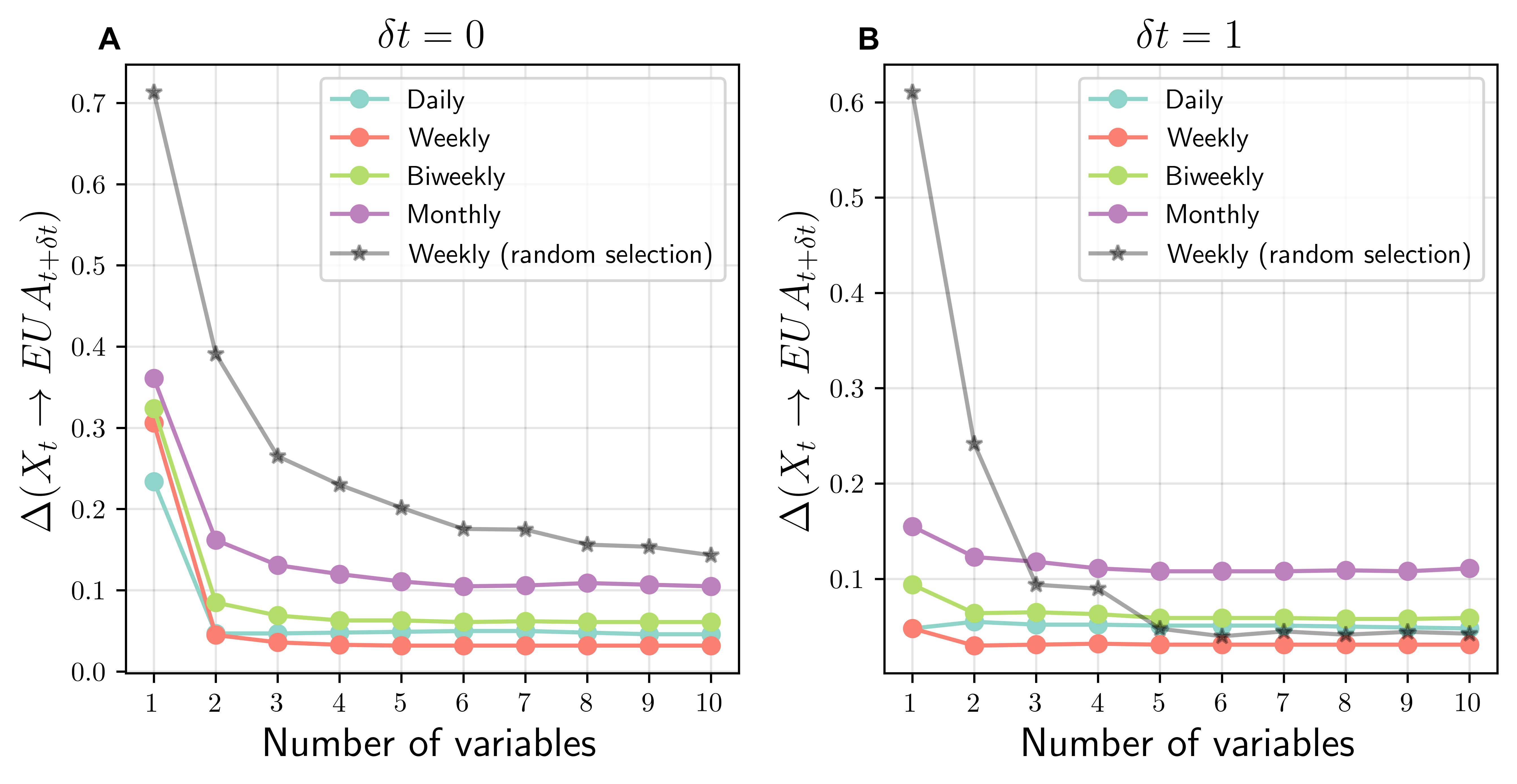}
            \includegraphics[width=0.98\columnwidth]{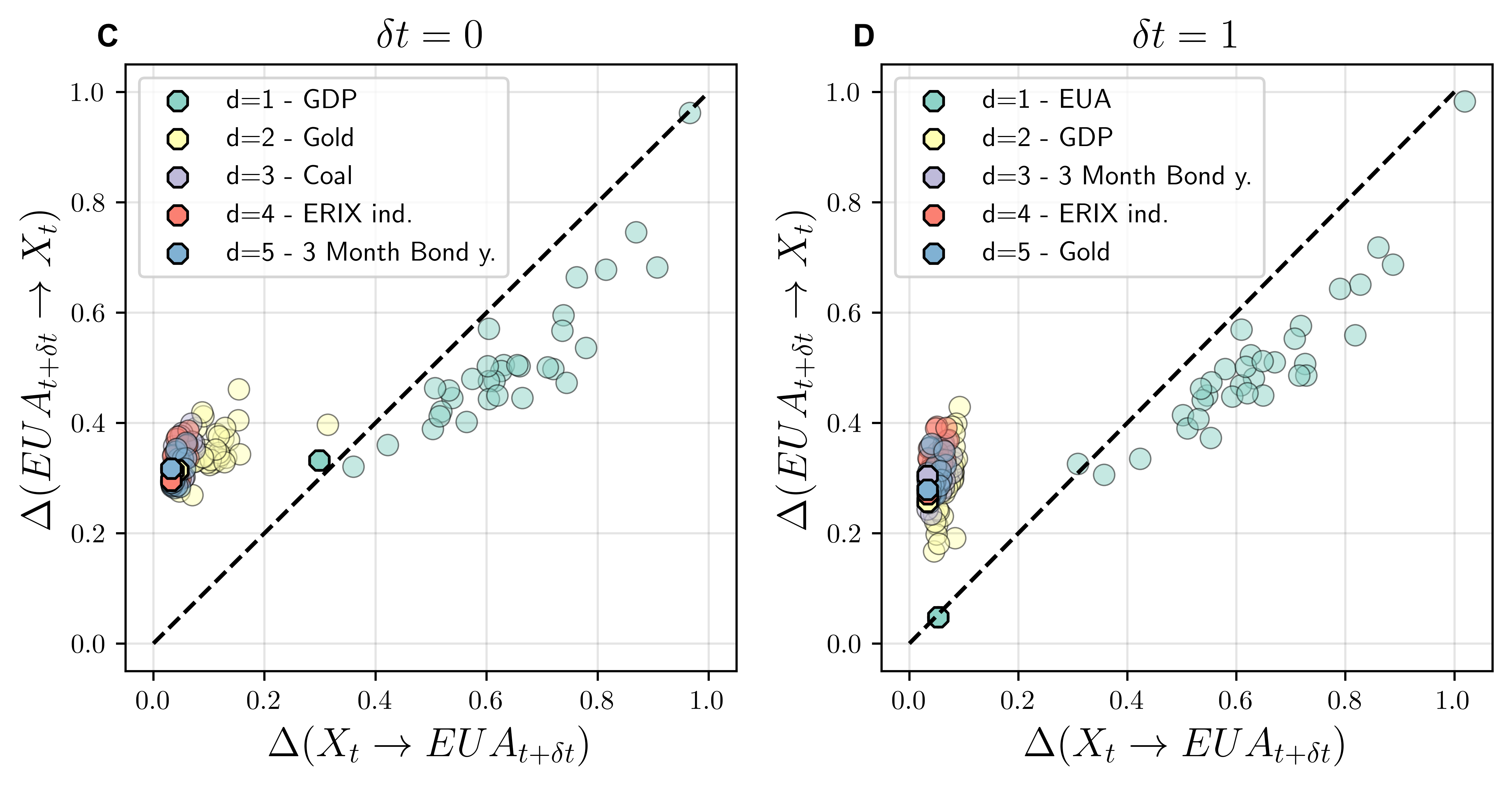}
        \caption{\textbf{Frequency identification and variable selection through the Information Imbalance.} The first rows plot the Information Imbalance $\Delta(X_t   \rightarrow EUA_t)$ from the predictor set to the EUA price for a growing number of variables in $X_t$ and for different data frequencies. The second row illustrates the greedy optimisation process to select the variables in $X_t$ for the most informative weekly frequency on the Information Imbalance plane.}
        \label{fig:II-TimeScale}
        \end{figure}
        Through the GP-based process of imputation and aggregation described in Section \ref{ssec:GPs}, we obtain 4 datasets at daily, weekly, biweekly, and monthly frequencies. Given these datasets, we use the Information Imbalance to identify the specific frequency at which the predictors are most informative about the EUA price. Specifically, for each frequency we perform the iterative greedy selection of variables  described in Section \ref{subsec:greedy_selection} aimed at minimising the Information Imbalance $\Delta (X_t \rightarrow EUA_{t+\delta t})$ from the set of predictors at time index $t$ ($X_t$), to the EUA price at time index $t + \delta t$ ($EUA_{t+\delta t}$). We perform such computations for a `nowcasting' scenario with $\delta t = 0$, and for a `forecasting' scenario with $\delta t = 1$, indicating one day, one week, two weeks or one month depending on the dataset.

        The resulting imbalances are graphed in Figure \ref{fig:II-TimeScale}. We find that, at all frequencies and for both time-lags $\delta t$, the information content of the predictor set does not improve substantially by adding more than $3$ variables, in agreement with the results already shown in Figure \ref{fig:II-TimeScale}. Furthermore, we find that for both time-lags, data on a weekly frequency contain the greatest information for predicting EUA price. This result indicates that smoothing the daily price oscillations on a weekly scale has a beneficial effect for easing the predictability of the EUA price, since such oscillations can hardly be interpreted using any of the considered features, but that smoothing over longer time scales erases important existing relationships and impairs predictive power. Given the greater information content of the weekly frequency, and also for convenience and brevity, the rest of the results in this section will be presented only for such frequency.
%
%   Subsection 5.2: Selection of predictor variables --------------------------------
    \subsection{Selection of predictor variables}
        \begin{figure*}[htp]
        \centering
            \includegraphics[width=0.98\textwidth]{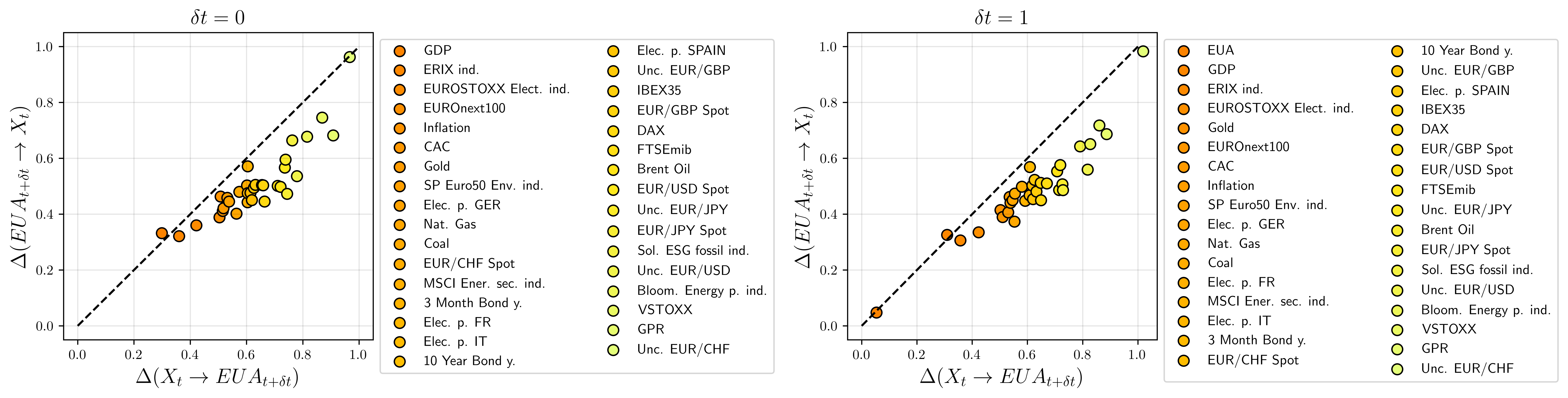}
        \caption{\textbf{Information Imbalance plane for nowcasting and forecasting EUA price.} The single variable information content toward the target EUA is presented in the same instant of time (left) and with 1 time lag. The results presented are at weekly frequency. The most informative variable are at the top of the legend in dark orange shades.}
        \label{fig:II-SingleVariablesWeekly}
        \end{figure*}
        This section presents an analysis of the predictors for nowcasting ($\delta t =0$) and forecasting ($\delta t =1$) using the Information Imbalance. The first result we analyse is the Information Imbalance calculated between the target variable (EUA) and each predictor presented in Table \ref{Table1:VariableDescription} taken individually (including Euro-area Inflation and GDP) for a dataset comprising both phases 3 and 4. Figure \ref{fig:II-SingleVariablesWeekly} presents the information content of all predictors with respect to the target variable. Observing the imbalances, we note that, in addition to GDP, the variables with higher predictive power are the ERIX index, EUROSTOXX Electricity index, and EUROnext100 index. Variables measuring uncertainty, like GPR and VSTOXX indices, on the contrary, have very low predictive power. In Figure \ref{fig:II-SingleVariablesWeekly}, we also present the results obtained by calculating the Information Imbalance in a forecasting framework, where we define the predictors set as $X_{t}$ and the target set as $\text{EUA}_{t+1}$. In particular, we observe that the lagged value of one week, $\text{EUA}_{t}$, is the most informative single variable. Even with a time lag, we see that GDP remains one of the most informative variables.

        Figure \ref{fig:II-TimeScale} presents the results obtained via the iterative greedy variable selection. It emerges that the most informative single variable is indeed GDP. This result confirms what emerges inspecting Figure \ref{fig:II-SingleVariablesWeekly}. The most informative subset of size 3 is GDP, Gold, and Coal prices. Given its nature as a composite long-term variable, GDP is undoubtedly a valid predictor for the long-term behaviour of economic and financial variables \citep{xu1996causality}, as is the case for EUA in our study. Therefore, this variable can be used to predict future trends in the price of our target variable. However, it is worth noting that the model  may have picked up a potential distortion effect. The trajectory of GDP resembles a  nearly continuous upward trend, mirroring the dynamics of the EUA price, which initially starts low and steadily rises in response to COVID and the energy crisis. Furthermore, Gold is considered a traditional safe-haven asset, suggesting that investors usually gravitate towards it
        during periods of economic uncertainty or market volatility. Gold's price movements often reflect shifts in market sentiment during times of economic instability or geopolitical uncertainty. Adding gold prices to a forecasting framework yields valuable insights into investor sentiment and anticipated market trends \citep{mohtasham2021deep}. Finally, the energy sector, particularly electricity generation, is heavily reliant on Coal prices. Considering the EU ETS encompasses various industries, notably power generation where coal is a major factor, changes in Coal prices serve as a significant indicator of broader energy market dynamics. Consequently, these shifts can impact the demand for, and pricing of, EU ETS allowances \citep{lovcha2022determinants}.

        Figure \ref{fig:II-TimeScale} also shows the results of the greedy approach in the presence of a time lag. Compared to the previous nowcasting setting, in this case, our methodology identifies very different informative subsets. In particular, now the most informative subset of 3 variables is composed of the lagged value of the target variable, GDP, and 3-month bond yield for the Euro-area. The relationship between the 3-month bond yield and EUA price may not be straightforward, but fluctuations in short-term interest rates, reflected in the 3-month bond yield, can indirectly influence economic conditions and investor actions, potentially impacting the demand for and pricing of EU ETS allowances \citep{chevallier2009carbon}.
%
%   Subsection 5.3: Prediction performances -----------------------------------------
    \subsection{Prediction performances}
        \begin{figure*}[htp]
            \centering
                \includegraphics[width=0.98\textwidth]{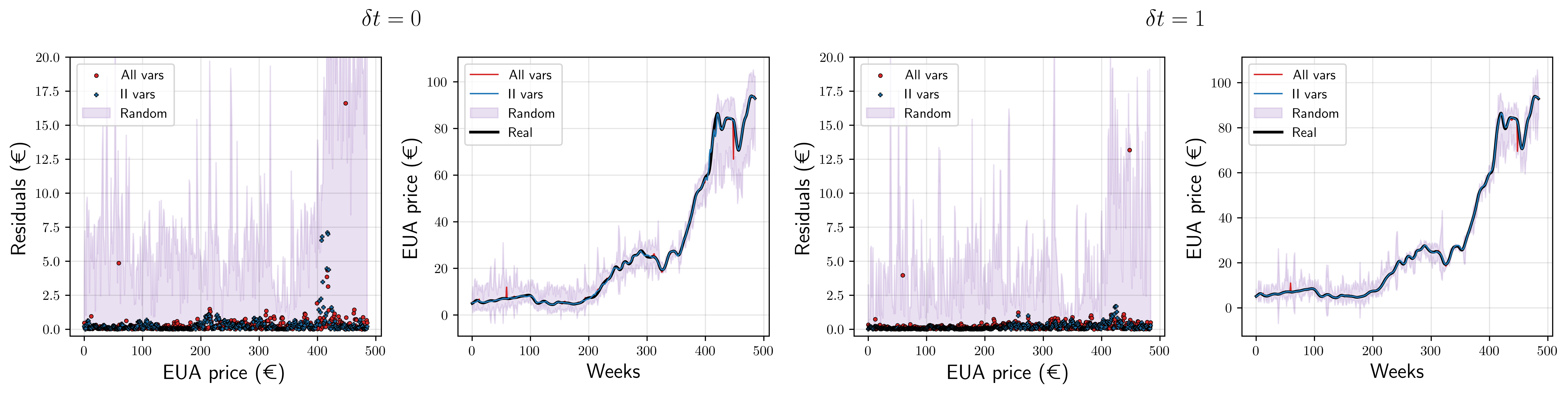}
            \caption{\textbf{Performance of a GP model build on: the 3 most informative variables; all variables; 3 randomly selected predictors (average over 10 replications).} For nowcasting (left) or forecasting (right) we report a scatter plot of the residuals and the predicted EUA time series together with the realised one.}
        \label{fig:Prediction}
        \end{figure*}
        We now verify that the information-driven and model-free variable selection performed in the last section translates into accurate prediction. As a benchmark model, we choose the same GP regression model described in Section \ref{sec:Methodology} and already used for imputation and aggregation purposes. Since the Information Imbalance indicates that the additional information content of predictors $X_t$ is marginal when compared to the one obtained with only 3 carefully selected variables, we compare the performance of a GP built using those 3 most informative variables, 3 randomly selected variables, and the full set of 33 variables in Nowcasting framework and 34 in Forecasting framework, respectively.

        Figure \ref{fig:Prediction} shows the result of such an experiment carried forward both for nowcasting ($\delta t = 0$) and forecasting ($\delta t = 1$). In agreement with the plateauing of information content evident from Figure \ref{fig:II-TimeScale}, the use of the full variable set in place of the 3 selected most informative variables does not further improve the performance. Remarkably, we observe that the use of only the most informative variables even leads to slightly more reliable and robust predictions compared to using all the variables in our dataset: the extra variables mostly act as noise leading to a deterioration of the predictive performance. The same effect is summarised in Table \ref{table:metrics}, which reports the mean squared error (MSE) computed on the cross-validation sets along with their level of uncertainty. For comparison, both the table and Figure \ref{fig:Prediction} also reports the performance obtained by averaging over 10 GP models on 3 exclusive and randomly selected variables, instead of the 3 variables selected using the Information Imbalance.
        \begin{table}[h]
        \setlength{\tabcolsep}{2pt}
        \centering
        \begin{tabular}{ccc}
            \multicolumn{3}{c}{Mean Squared Error} \\
            \hline
            \text{ } & $\delta t = 0$ & $\delta t = 1$ \\
            \hline
            \textit{Inf. Imb.} & \textbf{0.8}$\pm$0.3$\cdot10^{-3}$ & \textbf{0.1}$\pm$0.11$\cdot10^{-3}$ \\
            \textit{All} & 1.1$\pm$0.6$\cdot10^{-3}$ & 0.6$\pm$0.4$\cdot10^{-3}$ \\
            \textit{Rand.} & 79.5$\pm$21.2$\cdot10^{-3}$ & 34.8$\pm$6.8$\cdot10^{-3}$ \\
            \hline
        \end{tabular}
        \caption{\textbf{Prediction performance.} Mean squared error for nowcasting ($\delta t=0$) and forecasting ($\delta t=1$) GP models built using all predictors (\textit{All}), a set of 3 random predictors (\textit{Rand.}) and the set of 3 variables selected via the iterative greedy optimisation of the Information Imbalance (\textit{Inf. Imb.}). The GP model built on the 3 variables selected via Information Imbalance performs best.}
        \label{table:metrics}
        \end{table}
%
% Section 6: Conclusions ------------------------------------------------------------
\section{Conclusions}
\label{sec:Conclusions}
    Our study focuses on identifying exogenous drivers that influence the price of the EU ETS market and on providing an innovative methodology for mixed-frequency nowcasting and forecasting, all using a completely non-parametric approach. We have considered 33
    exogenous variables categorised into the following groups: 6 uncertainty-related variables, 8 commodities, 4 exchange rates, 6 energy indexes, 5 country-specific indexes, and 4 macroeconomic variables. Unlike conventional parametric econometric models, the Information Imbalance does not depend on any assumption on the model behind the time series and their relationships. Instead, it allows working with variables in their original form, ensuring results remain unaltered. This unquestionably stands out as a key advantage of the non-parametric approach employed, offering a greater degree of flexibility compared to other parametric analyses. For each of the last two EUA phases considered (Phase 3 and 4), we use the Information Imbalance to identify the most informative variables. We found that the most informative variables differ significantly between the two phases analysed. ERIX index, EUROSTOXX Electricity index, 3-month Bond yield were the top 3 most informative variables for Phase 3. For Phase 4, the most informative variables include Uncertainty on EUR/CHF exchange rate, Coal prices, and EUR/USD Spot rate. Consequently, we can conclude that during Phase 3, the most informative variables have a more fundamental nature, such as energy indices, commodities, and macroeconomic variables, while in Phase 4, financially oriented variables provide much informative content. This difference may be attributed to the impact of the COVID-19 pandemic and the energy crisis affecting the EUA price more significantly in Phase 4. These results were further supported by an iterative greedy selection of informative variables.
    In addition to empirical results, this article proposes a new methodology derived from the use of the Information Imbalance. Specifically, we have chosen to include Euro-area GDP and Inflation variables in our analysis at quarterly and monthly frequencies, respectively, with the intention of capturing longer-term economic cycle movements compared to variables observed in empirical findings. We utilised Gaussian processes to aggregate or impute the variables to ensure that they share the same temporal frequency. The informativeness of exogenous variables relative to the EUA target variable was measured via the Information Imbalance for each considered temporal scale, revealing the weekly scale as the most informative one. The high information content of the weekly dataset confirmed our choice to include longer-term macroeconomic variables, as in both the nowcasting framework ($\delta t=0$) and the forecasting framework ($\delta t=1$), the high informativeness of GDP and Inflation variables was evident. Our non-parametric analysis concludes with results obtained for nowcasting and forecasting predictions, once again achieved using Gaussian Processes. Predictions using all variables in our dataset and only the top 3 most informative variables were presented. As a benchmark, we also made predictions based on a random selection of variables. The results demonstrate that using only variables with high information content improves prediction performance, both for nowcasting as well as forecasting.
%
% Section: Replicating and supplementary materials ----------------------------------
\section*{Replicating and supplementary materials}
\label{sec:ReplicatingSupplementaryMaterials}
    \noindent
    Replicating and supplementary materials can be found at the following GitHub repository: \href{https://github.com/SaveChris/Inf-Imb-for-EUA23}{Information Imbalance for EUA 2023}
%
% Section: Acknowledgments ----------------------------------------------------------
\section*{Acknowledgments}
\label{sec:Acknowledgments}
    \noindent
    We acknowledge Antonio Di Noia (USI and ETH Zurich) for his contribution to the discussion regarding the utilisation of Gaussian Process regression for data imputation and aggregation.
    \\

    \noindent
    We extend our gratitude to Conor Hassan (QUT, Brisbane) for his attentive reading and valuable insights shared.
%
% Section: Funding -----------------------------------------------------------------
\section*{Funding}
\label{sec:Funding}
    \noindent
    The work by M. E. De Giuli has been supported by the Italian Minister of University and Research (MUR) through the following projects: "A geo-localized data framework for managing climate risks and designing policies to support sustainable investments" (No. 20229CWYXC) within the PRIN 2022 program; "Fin4Green - Finance for a Sustainable, Green and Resilient Society Quantitative approaches for a robust assessment and management of risks related to sustainable investing", (No.  2020B2AKFW-003) within the PRIN 2020 program. Her research was also supported by the European COST (Cooperation in Science and Technology through the project "Fintech and Artificial Intelligence in Finance –Towards transparent financial Industry", within the  COST Action CA19130.
    \\
    
    \noindent
    The work by A. Mira has been supported by Swiss National Science Foundation grant 200021\_208249.
% Section: CRediT authorship contribution statement --------------------------------
\section*{CRediT authorship contribution statement}
\label{AuthorContributions}
    \noindent
    \textbf{Cristiano Salvagnin:} Conceptualisation, Software, Validation, Formal analysis, Investigation, Data Curation, Writing - Original draft. \textbf{Aldo Glielmo:} Methodology, Software, Formal analysis, Writing - Review and Editing,  Supervision. \textbf{Maria Elena De Giuli:} Conceptualisation, Investigation, Writing - Original draft, Supervision, Funding acquisition. \textbf{Antonietta Mira:} Methodology, Writing - Review and Editing, Supervision, Funding acquisition.
%
% Section: Disclaimer ---------------------------------------------------------------
\section*{Disclaimer}
\label{sec:Disclaimer}
\noindent
    The views and opinions expressed in this paper are those of the Authors and do not necessarily reflect the official policy or position of Banca d’Italia.
%
% Section: Declaration of interest --------------------------------------------------
\section*{Declaration of interest}
\label{DeclarationOfInterest}
    \noindent
    Declaration of interest: none.
\newpage 
% References ------------------------------------------------------------------------
% \bibliographystyle{elsarticle-num-names}
\bibliographystyle{elsarticle-harv}\biboptions{authoryear}

\bibliography{IM-refs}
\end{document}